\documentclass[11pt,a4paper]{article}
\usepackage{cite}
\usepackage{amsmath}
\usepackage{amssymb}
\usepackage{mathtools}
\usepackage{stmaryrd}
\usepackage{float}
\usepackage{placeins}
\usepackage{hyperref}

\title{Enhanced posterior sampling via diffusion models for efficient metasurfaces inverse design}
\author{
	\textbf{Mathys Le Grand$^{1,2,*}$, \quad Pascal Urard$^{2}$,\quad Denis Rideau$^2$,}\\
		\textbf{\quad Loumi Tr\'emas$^2$, \quad Damien Maitre$^2$, \quad  Adam Fuchs$^2$}\\
		  \textbf{Louis-Henri Fernandez-Mouron$^2$,}\quad \textbf{R\'egis Orobtchouk$^1$} \\
	$^1$Institut des nanotechnologies de Lyon \quad $^2$STMicroelectronics \\
	\texttt{$^*$mathys.le-grand@insa-lyon.fr}}
\date{}

\begin{document}
	\maketitle
	\begin{abstract}
	 The inverse design of metasurfaces faces inherent challenges due to the nonlinear and highly complex relationship between geometric configurations and their electromagnetic behavior. Traditional optimization approaches often suffer from excessive computational demands and a tendency to converge to suboptimal solutions.
	 
	 This study presents a diffusion-based generative framework that incorporates a dedicated consistency constraint and advanced posterior sampling methods to ensure adherence to desired electromagnetic specifications. Through rigorous validation on small-scale metasurface configurations, the proposed approach demonstrates marked enhancements in both accuracy and reliability of the generated designs.
	 
	 Furthermore, we introduce a scalable methodology that extends inverse design capabilities to large-scale metasurfaces, validated for configurations of up to $98 \times 98$ nanopillars. Notably, this approach enables rapid design generation completed in minute by leveraging models trained on substantially smaller arrays ($23 \times 23$). These innovations establish a robust and efficient framework for high-precision metasurface inverse design.
	\end{abstract}

		\section{Introduction}
	
	Metasurfaces are subwavelength-patterned structures that enable precise manipulation of phase, amplitude, and polarization \cite{arbabi2015dielectric,yu2014flat}, with applications including dielectric metalenses \cite{lin2019topology,dilhan2021planar}, demultiplexers \cite{piggott2015inverse}, and beam shaping \cite{jafar2018adaptive,rideau2024approaches}. The inverse design problem entails identifying physical configurations that yield desired optical responses. However, the high dimensionality of the design space renders exhaustive exploration computationally prohibitive using traditional simulation techniques such as Finite Difference Time Domain (FDTD) \cite{gedney2011introduction}, Finite Element Method (FEM) \cite{rahman2013finite}, or spectral methods \cite{gaylord1985analysis}. Moreover, these methods face scalability limitations and often depend on periodic pattern assumptions.
	
	Deep learning-based surrogate models have been developed to accelerate simulations \cite{chen2022high}. However, the non-linear and non-convex relationship between structural parameters and optical properties introduces multiple local minima, complicating the optimization process. To mitigate these challenges, regularization techniques \cite{tikhonov1977solutions,vasin2014analysis} have been employed to reformulate ill-posed problems into well-posed ones. Another approach is direct structure prediction but despite its appeal, it often fails to converge due to the inherent one-to-many nature of the inverse problem \cite{liu2018training}.
	
	To address these challenges, generative methods such as Variational Autoencoders (VAEs) \cite{ma2019probabilistic}, Generative Adversarial Networks (GANs) \cite{so2019designing}, and Diffusion Models (DMs) \cite{zhang2023diffusion} have been introduced. Among these, diffusion models, though less explored for metasurface design, exhibit superior sample quality and training stability compared to GANs \cite{dhariwal2021diffusion}. 
	
	Diffusion models, a type of generative model, have been recently applied in material science to design structures such as porous media \cite{park2024inverse}, stress distribution \cite{bastek2023inverse}, and effective doping of superconductors \cite{zhong2024high}. These studies have noted that DMs outperform GANs in the generation of structures. The first example of using DMs for photonic devices is presented in \cite{zhang2023diffusion}. This work compares DMs against GANs and improved GANs like WassersteinGAN \cite{salimans2016improved}, concluding that DMs are more accurate for generating structure parameters and more stable during training, making them more efficient to work with.
	
	This paper investigates the application of DMs to the inverse design of metasurfaces, building on the ancestral sampling approach introduced in \cite{zhang2023diffusion}. Despite promising results, ancestral sampling often fails to satisfy design constraints. To address this, we introduce a consistency term in the loss function and explore posterior sampling techniques to improve compliance to input conditions. We first demonstrate the performance gains of diffusion models on small metasurfaces and then present a method to scale the inverse design to larger metasurfaces.
	
	The code developed for this study is open-source at \cite{le_grand_2026_18148500}.
	
	\section{Physical System}
	
	We address the inverse design of metasurfaces for a beam-shaping problem, a methodology generalizable to various photonic and material science applications. Metasurfaces are engineered structures composed of subwavelength elements that enable manipulation of electromagnetic waves beyond the capabilities of natural materials. Their design requires a thorough understanding of the interaction between electromagnetic fields and the metasurface, which can be categorized into near field and far field effects.

	The near field refers to the region within a few wavelengths of the metasurface, where electromagnetic fields are highly localized and exhibit strong spatial variations. This region is dominated by evanescent waves, which decay exponentially with distance. Near field interactions are critical for applications such as sensing, imaging, and energy harvesting, where precise control of the local electromagnetic environment is essential. In this work, the near field serves as an intermediate result necessary for computing the far field.

	The far field, in contrast, refers to the region far from the metasurface, where the distance is much greater than the wavelength of the interacting waves. Here, electromagnetic fields propagate as plane waves with smoother spatial variations. The far field behavior determines how the metasurface manipulates the direction, phase, and polarization of propagating waves, which is crucial for applications such as beam shaping, holography, and communication systems.

	To compute the electromagnetic field interaction with the metasurface, we first obtain the near field using the FDTD method \cite{gedney2011introduction}. The near field is then propagated to the far field using the Fraunhofer diffraction approximation, which applies a Fourier transform to the near field data \cite{born2013principles,umashankar1982novel}.

	The metasurface architecture is engineered to modulate both near and far field electromagnetic interactions. Through precise manipulation of the geometry, dimensions, and spatial configuration of subwavelength constituents, tailored electromagnetic responses are realized. This work specifically investigates the optimization of far field power distribution.
	
	The optimized structure comprises a regular grid of dielectric pillars, with a center-to-center spacing of $\lambda/2$. The incident electromagnetic field is a normally incident monochromatic plane wave at wavelength $\lambda$.
	
	\section{Related works}
	
	Prior research has explored the application of diffusion models to the inverse design of metasurface structures \cite{zhang2023diffusion,zhang2024addressing,hen2025inverse,seo2025physics}. Table~\ref{table:rela_work} provides a comparative overview of these studies across various topics.
	
	\begin{table*}[h]
		\centering
		\caption{Comparison of topics addressed in related works on diffusion models for metasurface inverse design.}
		\begin{tabular}{|p{2.5cm}|p{1.7cm}|p{1.7cm}|p{2cm}|p{2cm}|p{2cm}|}
			\hline
			Topic &\cite{zhang2023diffusion}&\cite{zhang2024addressing}&\cite{hen2025inverse}&\cite{seo2025physics}& Our work  \\
			\hline
			Figure of Merit & S-parameter & S-parameter  &Far Field power distribution for beam shaping &Far Field power distribution for color routing & Far Field power distribution for beam shaping\\ 
			\hline
			
			Structure &Freeform&Freeform  &Freeform&Freeform & Pillars \\ 
			\hline
			Posterior Sampling& \textbackslash  & \textbackslash &-Raw  & -Raw &-Raw \newline -Monte Carlo \cite{chung2022diffusion} \\ 
			\hline
			Constraint Posterior Sampling&\textbackslash   &\textbackslash  & \textbackslash    &\textbackslash   & Spherical Gaussian constraint\cite{yang2024guidance} \\ 
			\hline
			Gradient computation related to Posterior Sampling&\textbackslash   &\textbackslash  & Differentiable RCWA  \cite{kim2023torcwa}& Adjoint method \cite{giles2000introduction} with FDTD from MEEP \cite{oskooi2010meep}& Differentiable surrogate trained on FDTD samples \cite{rideau2024approaches} \\ 
			\hline
			
			Consistency loss& \textbackslash&\textbackslash& \textbackslash  &\textbackslash & \checkmark \\ 
			\hline
			Scaling (Higher Degrees of Freedom) & \textbackslash&\textbackslash& \textbackslash &\textbackslash &\checkmark \\ 
			\hline
			
		\end{tabular}
		
		\label{table:rela_work}
	\end{table*}
	
	In this work, we opted to utilize a surrogate network due to its computational speed, differentiability, and scalability properties that are significantly constrained in rigorous methods such as FDTD or RCWA \cite{kim2023torcwa}. However, this approach introduces a dependency on the surrogate's precision, necessitating the development of a highly performant surrogate model as a critical component of this study \cite{rideau2024approaches}.
	
\begin{figure*}[h]
		\centering
		\includegraphics[scale=0.55]{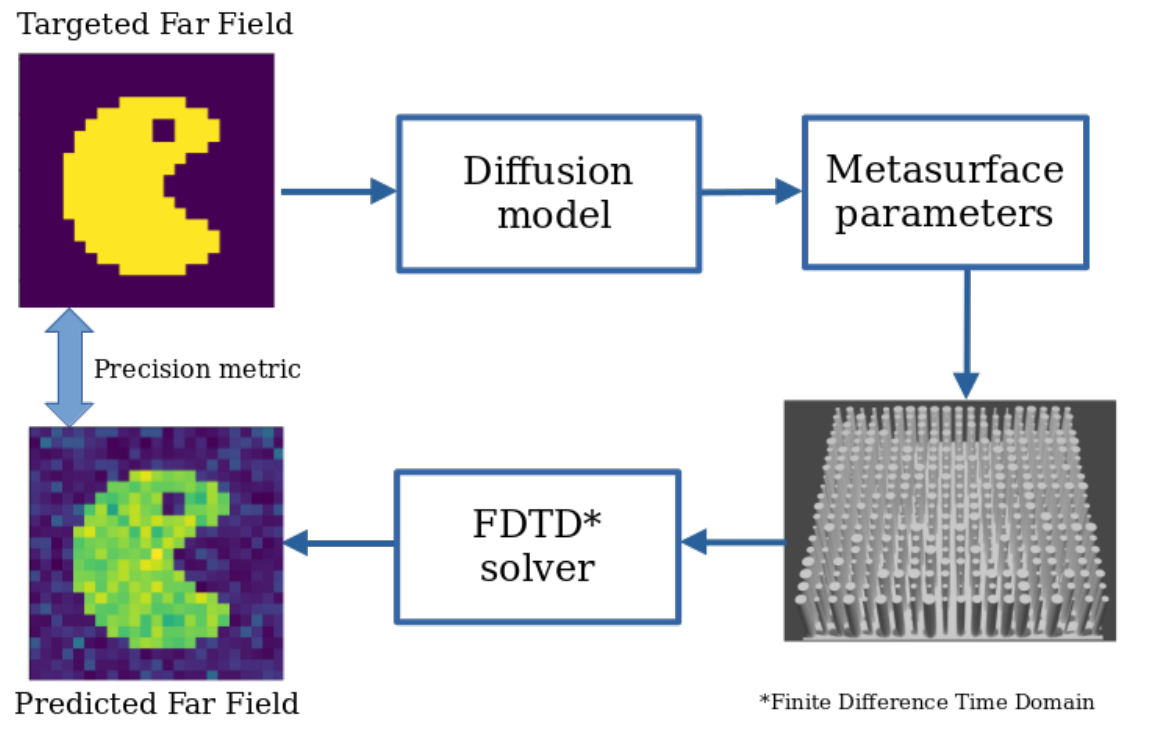}
		\caption{Verification process of the inverse design flow using diffusion models.}
		\label{fig:verif_process}
	\end{figure*}

	These methods are applied to the inverse design of a beam-shaping metasurface composed of pillars with varying radii, targeting a specified far field amplitude distribution. Optimization results are validated by comparing the simulated far field response to the target using the $R^2$ metric. Given the nonlinear nature of the problem, the $R^2$ metric ranges over $(-\infty, 1]$, where values closer to 1 indicate better agreement. The evaluation workflow is shown in Figure~\ref{fig:verif_process}.

	\section{Reference methods} \label{sec:ref}

	In the introduction, several generative methods such as VAEs \cite{ma2019probabilistic} and  GANs \cite{so2019designing} were discussed. However, these approaches are excluded from comparative analysis due to their demonstrated inadequacy in far field metasurface inverse design, as evidenced in \cite{hen_yosef_raviv_giryes_scheuer_2025}.
	Instead, this study evaluates a diffusion model incorporating enhanced posterior sampling and training techniques, benchmarked against a phase retrieval and look-up-table method, as well as gradient descent utilizing a surrogate network.
	\subsection{Phase Retrieval \& Look-Up Table}
	The primary challenge in metasurface design lies in the fact that only the far field amplitude is available, making the direct retrieval of the near field via an inverse Fourier Transform infeasible. To address this, a phase retrieval algorithm \cite{jaganathan2016phase,wang2017hybrid} is employed to estimate the near field phase profile. This phase profile is then used as input to a bijective function that maps local phase shifts to meta-atom parameters. This method, which combines phase retrieval with a look-up table, operates under two key assumptions: (1) the transmission of a meta-atom is independent of its parameter, and (2) there are no interactions among the meta-atoms within the metasurface. These methods are ineffective for metasurfaces exhibiting strong inter-meta-atom coupling~\cite{isnard2024advancing}, as well as for stacked metasurface layers with significant inter- and intra-layer interactions. In contrast, inverse design methods based DMs do not rely on such stringent approximations and can seamlessly scale to accommodate a large number of parameters per meta-atom, making them a more robust alternative.
	\subsection{Gradient descent}
	
	Gradient Descent (GD) is a fundamental optimization algorithm widely used in machine learning to minimize objective functions. The goal is to iteratively adjust parameters to minimize a loss function, which quantifies the discrepancy between model predictions and target data. The parameter update rule is expressed as:
	
	\begin{equation}
		\theta = \theta - \alpha \nabla_{\theta} L(\theta),
	\end{equation}
	
	where $\alpha$ is the learning rate, a hyperparameter controlling the step size, and $\nabla_{\theta} L(\theta)$ is the gradient of the loss function $L(\theta)$ with respect to the parameters $\theta$.
	
	In the context of metasurface design, the parameters to optimize are the metasurface parameters. The loss function is defined as the distance between the target far field amplitude and the simulated far field derived from the metasurface parameters. To make this optimization computationally efficient, a surrogate solver based on a neural network is used instead of rigorous solvers like FDTD. The surrogate solver significantly accelerates simulations and, being fully differentiable, enables efficient gradient computation.
	
	Modern deep learning frameworks, such as Jax and PyTorch, leverage automatic differentiation to compute gradients efficiently. Automatic differentiation applies the chain rule to compute derivatives of complex functions by breaking them into simpler operations. This process is managed by the framework’s computational graph, which records the sequence of operations applied to the inputs.
	
	Once the computational graph is constructed, backpropagation is performed by traversing the graph in reverse to compute the gradients of the loss function with respect to each parameter. These gradients are then used to update the parameters using optimization algorithms such as Stochastic Gradient Descent (SGD) \cite{amari1993backpropagation}, ADAM \cite{adam2014method}, or RMSprop. In this study we use GD with ADAM.
	
	The development of automatic differentiation was initially aimed at facilitating the training of complex neural networks, but it has since been extended to differentiable solvers \cite{kim2023torcwa,ponomareva2025torchgdm}. These solvers are now indispensable tools for efficient gradient-based optimization in metasurface design and other scientific applications.

	\section{Diffusion models}
	
	\subsection{Basic Concept}
	Diffusion models aim to model the data generation process as a gradual transformation of simple noise into complex data. This is achieved through a series of incremental steps, each refining the data slightly. The process is conceptualized as a reverse diffusion process, where noise is iteratively denoised to generate realistic samples \cite{sohl2015deep}.
	
	\subsection{Forward and Reverse Processes}
	Diffusion models consist of two main processes:
	\begin{itemize}
		\item \textbf{Forward Process:} Gaussian noise is added to the data over a series of steps. Starting from the original data, noise is incrementally added until the data is completely corrupted and resembles pure noise \cite{ho2020denoising}.
		\item \textbf{Reverse Process:} This process reconstructs the original data from its noisy version by learning a series of denoising steps. These steps gradually remove the noise until the data is restored \cite{song2020score}. The reverse process introduced in \cite{song2020score} is referred to as ancestral sampling in this paper, and its performance is compared to posterior sampling \cite{chung2022diffusion}.
	\end{itemize}
	
	\subsection{Mathematical Formulation}
	Let $x_0$ represent the original data and $x_T \sim \mathcal{N}(0, \mathbf{I})$ represent the completely noisy data. The forward process is defined by Gaussian transitions:
	
	\begin{equation}
		q(x_t | x_{t-1}) = \mathcal{N}(x_t; \sqrt{1 - \beta_t} x_{t-1}, \beta_t \mathbf{I}),
	\end{equation}
	
	where $\beta_t$ is a variance schedule controlling the noise added at each step $t$. We define $\alpha_t$ as :
	\begin{equation}
		\alpha_t = \prod_{s=1}^{t}(1 - \beta_s),
		\label{eq:alpha} 
	\end{equation}
	
	Hence, the transition for any step $t \in [0, T]$ from the original data $x_0$ can be expressed as:
	
	\begin{equation}
		q(x_t | x_0) = \mathcal{N}(x_t; \sqrt{\alpha_t} x_0, (1 - \alpha_t) \mathbf{I}),
	\end{equation}
	
	where $\mathbf{I}$ is the identity matrix. Hence, the noisy sample $x_t$ can be written as:
	
	\begin{equation}
		x_t = \sqrt{\alpha_t} x_0 + \sqrt{1 - \alpha_t} \epsilon, \quad \epsilon \sim \mathcal{N}(0, \mathbf{I}).
	\end{equation}
	
	The reverse process learns the conditional distribution $p_\theta(x_{t-1} | x_t)$, parameterized as:
	
	\begin{equation}
		p_\theta(x_{t-1} | x_t) = \mathcal{N}(x_{t-1}; \mu_\theta(x_t, t), \sigma_t^2 \mathbf{I}),
	\end{equation}
	
	where the predicted mean $\mu_\theta(x_t, t)$ is derived using the posterior mean $x_{0|t}$, computed as:
	
	\begin{equation}
		x_{0|t} = \mathbb{E}[x_0 | x_t] = \frac{x_t - (1 - \alpha_t) \nabla_{x_t} \log p(x_t)}{\sqrt{\alpha_t}},
		\label{posterior_mean_score}
	\end{equation}
	
	where $\nabla_{x_t} \log p(x_t)$, the score, is approximated by a neural network $\epsilon_\theta(x_t, t)$:
	
	\begin{equation}
		x_{0|t} \approx \frac{x_t - \sqrt{1 - \alpha_t} \epsilon_\theta(x_t, t)}{\sqrt{\alpha_t}}.
		\label{posterior_mean}
	\end{equation}
	
	The predicted mean is then given by:
	
	\begin{equation}
		\mu_\theta(x_t, t) = \sqrt{\alpha_{t-1}} x_{0|t} + \sqrt{1 - \alpha_{t-1} - \sigma_t^2} \epsilon_\theta(x_t, t).
		\label{predicted_mean}
	\end{equation}
	
	Finally, the reverse process update rule is:
	
	\begin{equation}
		x_{t-1} = \mu_\theta(x_t, t) + \sigma_t \epsilon,
	\end{equation}
	
	where $\sigma_t = \eta \sqrt{1 - \frac{\alpha_{t-1}}{1 - \alpha_t}} \sqrt{1 - \frac{\alpha_t}{\alpha_{t-1}}}$, with $\eta \in [0,1]$ being a parameter controlling the stochasticity of the sampling. For $\eta = 0$, the sampling process becomes deterministic.
	
	\subsection{Training}
	Training diffusion models involves optimizing the parameters $\theta$ to minimize the difference between the true and learned reverse processes. This is typically achieved using variational inference techniques by minimizing a variational bound on the negative log-likelihood of the data \cite{kingma2021variational}.
	
	\subsection{Conditional Diffusion Models}
	Conditional diffusion models extend the basic framework by generating data samples conditioned on specific input information or context. These models guide the data generation process using transformations influenced by the conditional input, enabling the generation of targeted outputs. For example, in this work, the condition corresponds to the far field shape.
	
	Eq. \eqref{predicted_mean} is modified for conditional diffusion models as:
	
	\begin{equation}
		\mu_\theta(x_t, \mathbf{c}, t) = \sqrt{\alpha_{t-1}} x_{0|t} + \sqrt{1 - \alpha_{t-1} - \sigma_t^2} \epsilon_\theta(x_t, \mathbf{c}, t),
		\label{conditional_predicted_mean}
	\end{equation}
	
	where $\mathbf{c}$ denotes the condition vector. Unlike \cite{zhang2023diffusion}, conditioning alone was insufficient to achieve satisfactory results. To enhance performance, modifications were introduced in both training and sampling. Specifically, a consistency loss term was incorporated to provide additional guidance. A detailed comparison of these methods is presented in Section~\ref{sec:comp_res}.

	\subsection{Architecture}
	A core architectural requirement is scalability, enabling training on images of a specific size while supporting generation at larger dimensions. This was achieved using a fully convolutional network architecture \cite{long2015fully}, selected for its inherent scalability. Future work may explore self-attention mechanisms to further enhance performance \cite{zhao2020exploring}.
	
	The architecture employs skip connections and residual blocks to improve training depth \cite{ronneberger2015u,he2016deep}. For time embedding, a sinusoidal approach with Group Normalization was implemented. When using constant or sinusoidal time embeddings that introduce channel-wise constants, Batch Normalization should be avoided to prevent vanishing timestep embeddings \cite{bach2015batch,kim2024disappearance}. Alternative normalization techniques, such as Group Normalization \cite{wu2018group}, Layer Normalization \cite{ba2016layer}, or advanced embedding methods like positional timestep embedding \cite{kim2024disappearance}, are recommended.
	
	\section{Training Improvements}
	
	In this section, two methods aimed at improving the training procedure of DMs are presented. While these methods influence the training metrics, such metrics alone are insufficient to determine the optimal training approach. Instead, the selection criterion is based on the final performance metric computed on the far field response obtained after inverse design and rigorous simulation using FDTD. Throughout this paper, the term final metric refers to the evaluation performed on the far field resulting from inverse design followed by rigorous simulation. The following conclusions are drawn using the improved sampling techniques introduced in the subsequent section. The ancestral sampling method proposed by \cite{ho2020denoising} is ineffective for the inverse design problem addressed here.

	\subsection{Noising Schedule} \label{Sec:schedule}

	As demonstrated by \cite{nichol2021improved}, the choice of noise schedule significantly impacts the performance of diffusion models across different applications. To optimize far field precision, we introduced three distinct schedules for the noise variance $\beta(t)$: \textit{Linear}, \textit{Quadratic}, and \textit{Sigmoid}. The function $\beta(t)$ is monotonically increasing, satisfying $\beta(0) = \beta_{\text{start}}$ and $\beta(T) = \beta_{\text{end}}$. Detailed definitions of these schedules are provided in Appendix \ref{App:schedule}.
	
	Notably, an improved training metric does not guarantee superior inverse design performance. While the training metric may appear high, the final result evaluated through the verification process depicted in Figure~\ref{fig:verif_process} can remain unsatisfactory. Thus, computing the final metric is essential for reliably assessing model performance and selecting the optimal schedule. The correspondence between the training metric and the final result metric is further analyzed in Appendix \ref{App:schedule}.

	\subsection{Consistency Loss}
	
	Incorporating a consistency loss during the training of DMs can significantly enhance the quality and reliability of the generated data. DMs, which transform data from a simple prior distribution to a complex target distribution, benefit from additional constraints that ensure coherence and consistency throughout the diffusion process.

	The purpose of consistency loss is to enforce desirable properties on the intermediate states of the data during the diffusion process. It preserves the integrity and continuity of the data by penalizing deviations from expected patterns or structures. This loss is particularly beneficial in conditional generation tasks, ensuring that the outputs at each step consistently reflect the conditioning information.
	
	To implement consistency loss, a metric is defined to quantify the discrepancy between the conditioning information and the predicted condition derived from the posterior mean during the diffusion process. This metric may measure structural similarity, feature consistency, or adherence to the conditioning information. For instance, in image generation tasks, consistency loss could involve comparing pixel-wise differences or feature maps across consecutive diffusion steps to ensure smooth transitions and coherent structures. In this work, we utilize the posterior mean $x_{0|t}$ derived from Eq. \eqref{posterior_mean} at each step $t$, introducing an additional loss term.
	
	Using Tweedie's formula \cite{efron2011tweedie}, the posterior mean $x_{0|t}$ is fed into a surrogate simulator $S_\phi$, which takes metasurface parameters as input and outputs the near electromagnetic field. This field is then propagated using a Fourier transform to compute the far field (FF) for the Finite-Difference Time-Domain (FDTD) method. A surrogate is employed here to replace the computationally expensive FDTD simulation. During training, the DM is penalized when $S_\phi(x_{0|t})$ deviates significantly from the input condition $c$. The details of the surrogate simulator are addressed in Appendix \ref{App:surrogate}.
	
	The consistency loss is formally expressed as:
	
	\begin{equation}
		\mathcal{L} = \mathcal{L}_{diff}(\epsilon_\theta(x_t, c, t), \epsilon) + \\ 
		\gamma_t \cdot \mathcal{L}_{consistency}(S_\phi(x_{0|t}), c),
		\label{weighted_consistency_loss}
	\end{equation}
	
	where $x_{0|t}$ is obtained from Eq.~\eqref{posterior_mean}, $\mathcal{L}_{diff}$ represents the diffusion loss, and $\mathcal{L}_{consistency}$ measures the discrepancy between the surrogate output $S_\phi(x_{0|t})$ and the conditioning information $c$.
	
\begin{figure}[h]
		\centering
		\includegraphics[scale=0.65]{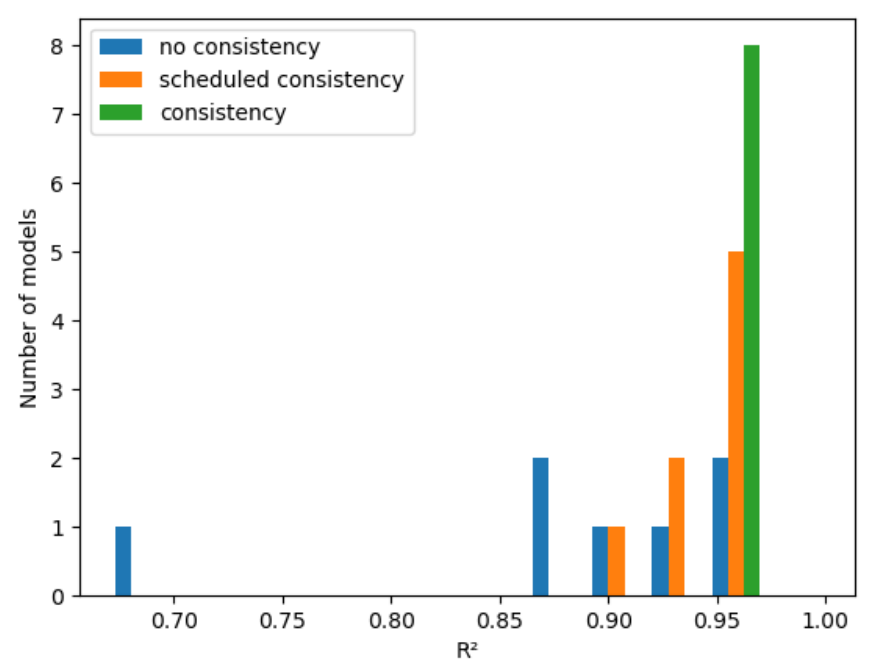}
		\caption{Final metric results for the three type of consistencies (no consistency loss, consistency loss and scheduled consistency loss) used during training. The results exhibit greater concentration near the optimal value when consistency loss is applied. }
		\label{fig:consistency_loss}
	\end{figure}

	The contribution of the consistency loss to the overall loss can be modulated in two ways:
	
	\begin{itemize}
		\item \textbf{Uniform weighting:} The consistency loss is included uniformly across all time steps, with $\gamma_t = 1$, as illustrated by the ``consistency'' data in Figure~\ref{fig:consistency_loss}.
		\item \textbf{Dynamic weighting:} The consistency loss is weighted by $\gamma_t=\alpha_t$, which varies with the diffusion step, as shown by the ``scheduled consistency'' data in Figure~\ref{fig:consistency_loss}. Here, its influence increases as $t$ approaches zero and the data coefficient $\alpha_t$ approaches one, emphasizing the consistency loss in the final steps.
	\end{itemize}
	
	While the concept of enforcing forward-backward consistency is well-established, e.g., in cycle consistency for GANs \cite{zhu2017unpaired}, the proposed consistency loss differs from the ``Consistency Models'' introduced in \cite{song2023consistency}. The latter focuses on reducing the number of sampling steps without imposing additional constraints for conditional generation.
	
	The incorporation of consistency loss significantly reduces the sensitivity of diffusion models to hyperparameter variations. As depicted in Figure~\ref{fig:consistency_loss}, the distribution of final metric results with consistency loss is markedly more concentrated around optimal outcomes, in contrast to results obtained without consistency loss, which fall below the performance of state-of-the-art methods discussed in Section~\ref{sec:ref}.
	
	When employing dynamic weighting of the consistency loss, the result distribution exhibits greater variability compared to uniform weighting, which consistently yields only the best-performing diffusion models. Thus, the introduction of consistency loss effectively mitigates the impact of hyperparameter fluctuations, simplifying the optimization process for diffusion model hyperparameters.

	\section{Improved sampling} \label{sec:improved_sampling}
	Results presented in this section are derived from the best-performing model selected through training improvement techniques. Model performance was consistent across the three enhanced sampling methods introduced herein.
	
	\subsection{Posterior Sampling}
	
	Incorporating additional information about the condition during the training phase proved insufficient to achieve satisfactory results. To address this, a guidance term is introduced during the sampling phase, leading to a new sampling method referred to as posterior sampling. Similar to the consistency loss, this sampling strategy leverages Tweedie's formula \cite{efron2011tweedie} to compute the guidance term from the posterior mean $x_{0|t}$. However, instead of using the score $\nabla_{x_t}\log p(x_t)$ in Eq. \eqref{posterior_mean_score}, the condition is explicitly incorporated by computing $\nabla_{x_t}\log p(x_t|c)$. Using Bayes' rule, this can be expressed as:
	
	\begin{equation}
		\nabla_{x_t}\log p(x_t|c) = \nabla_{x_t}\log p(c|x_t) + \nabla_{x_t}\log p(x_t).
	\end{equation}
	
	The new term $\nabla_{x_t}\log p(c|x_t)$ must now be computed. To make this term tractable, the Jensen approximation is employed. For a random variable $x$ following a distribution $p(x)$ and a function $f$, the approximation states:
	
	\begin{equation}
		\mathbb{E}_{x \sim p(x)}[f(x)] \approx f(\mathbb{E}_{x \sim p(x)}[x]).
	\end{equation}
	
	Applying this approximation to posterior sampling results in:
	
	\begin{equation}
		p(c|x_t) \approx p(c|x_{0|t}).
	\end{equation}
	
	This term can now be evaluated using an electromagnetic simulator, approximated by the neural network $S_\phi$, such that $S_\phi(x_{0|t}) = c$. Using this approximation, the gradient term also dubbed guidance term becomes:
	
	\begin{equation}
		\nabla_{x_t}\log p(c|x_t) \approx \frac{-1}{\sqrt{1-\alpha_t}} \nabla_{x_t}||c - S_\phi(x_{0|t})||^2.
	\end{equation}
	
\begin{figure}[h]
		\centering
		\includegraphics[scale=0.5]{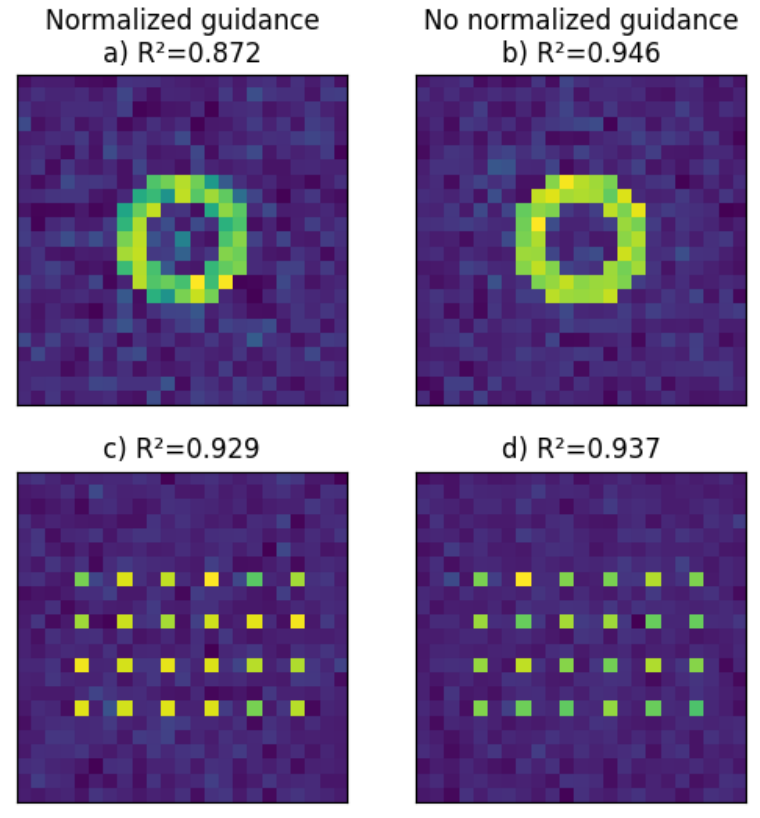}
		\caption{$|$Far field$|$ after simulation from inverse-designed metasurface parameters using posterior sampling with 1k steps. Normalized guidance is defined as $q_t = \frac{1}{\|c - S_\phi(x_{0|t})\|_2}$, whereas non-normalized guidance corresponds to $q_t = 1$.}
		\label{fig:dps_results}
	\end{figure}

	The update rule from Eq. \eqref{predicted_mean} is then modified as:
	
	\begin{equation}
		\label{predicted_mean_posterior_sampling}
		\mu_\theta^{ps}(x_t, c, t) = \mu_\theta(x_t, c, t) 
		- \frac{1}{\sqrt{1-\alpha_t}} \nabla_{x_t}||c - S_\phi(x_{0|t})||^2.
	\end{equation}
	
	\begin{equation}
		\label{eq:predicted_mean_posterior_sampling_q}
		\mu_\theta^{ps}(x_t, c, t) = \mu_\theta(x_t, c, t) \\
		- q_t\nabla_{x_t}||c - S_\phi(x_{0|t})||^2.
	\end{equation}
	With $q_t =\frac{1}{\sqrt{1-\alpha_t}}$.
	\cite{chung2022diffusion} argues that experimental results exhibit greater stability when using the weighting $q_t = \frac{\text{cte}}{\|c - S_\phi(x_{0|t})\|_2}$. This observation is supported by \cite{song2023loss}, which demonstrates through a Gaussian example that the guidance term tends to be overestimated at the start of the generation when $\sigma_t$ is large, and underestimated near the end when $\sigma_t$ is small. However, as demonstrated in Figure~\ref{fig:dps_results}, normalizing the gradient term slightly reduces precision on the final metric for $23 \times 23$ pillar metasurfaces. Conversely, when scaling to larger metasurfaces and increasing the dimensionality of the optimization problem, gradient normalization improves performance, as evidenced in Section \ref{sec:scaling}.
	Posterior sampling yields promising results, as shown in Figure \ref{fig:dps_results}. The shape specified as the condition is now recognizable in the verification simulation. Ancestral sampling as proposed in \cite{ho2020denoising} failed to generate shapes that closely resemble the conditioning input. Furthermore, it is possible to combine the consistency loss approach with posterior sampling, as discussed in Section \ref{sec:comp_res} on comparative results.
	
	\subsection{Monte Carlo Posterior Sampling}
	
	Monte Carlo posterior sampling (PS MC) \cite{song2023loss } is a variation of the posterior sampling method introduced in the previous section. Instead of relying solely on $x_{0|t}$ to compute the guidance term $\nabla_{x_t} \|c - S_\phi(x_{0|t})\|_2^2$, $x_{0|t}$ is treated as the mean of a Gaussian distribution, $x_{i,0|t} \sim \mathcal{N}(x_{0|t}, r_t)$, where the hyperparameter $r_t$ represents the guidance schedule. This method aims to mitigate the guidance value overshoot and undershoot discussed in the previous section. The new guidance term is then expressed as:
	
	\begin{equation}
		\nabla_{x_t}\log\left(\frac{1}{N}\sum_{i=1}^{N}\exp(-\|c - S_\phi(x_{0|t}^i)\|_2^2)\right).
	\end{equation}
	
	The corresponding update rule becomes:
	
	\begin{equation}
		\mu_\theta^{mc}(x_t, c, t) = \mu_\theta(x_t, c, t) + 
		\nabla_{x_t}\log\left(\frac{1}{N}\sum_{i=1}^{N}\exp(-\|c - S_\phi(x_{0|t}^i)\|_2^2)\right).
	\end{equation}
	
	As highlighted in \cite{song2023loss}, using a simple example of a Gaussian mixture, Monte Carlo sampling provides a more accurate scale for the guidance term compared to standard posterior sampling. This improvement allows for a more precise computation of the guidance term. 
	
\begin{figure}[h]
		\centering
		\includegraphics[scale=0.6]{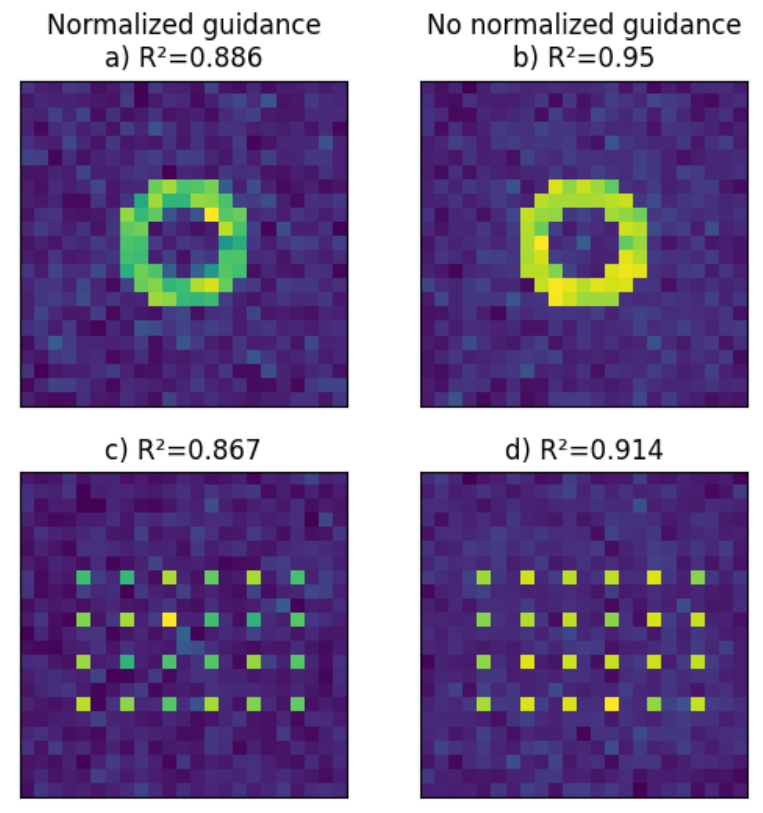}
		\caption{$|$Far field$|$ obtained from inverse-designed metasurface parameters using Monte Carlo posterior sampling with $1,000$ steps and $N=5$. The normalized guidance is defined as $q_t = \frac{1}{\|c - S_\phi(x_{0|t})\|_2}$, while non-normalized guidance corresponds to $q_t = 1$.}
		\label{fig:dps_mc_results}
	\end{figure}

	As illustrated in Figures~\ref{fig:dps_results} and~\ref{fig:dps_mc_results}, standard posterior sampling and Monte Carlo posterior sampling exhibit comparable performance. Empirically, the guidance schedule $r_t = (1 - \alpha_t)^2$ yields the best results. However, Monte Carlo posterior sampling incurs a substantial computational overhead: for $N$ samples, both generation time and memory usage scale linearly with $N$. In subsequent figures, the number of Monte Carlo samples for posterior sampling, denoted as $N$, will be represented as $MC_N$.
	
	\subsection{Spherical Gaussian Constrained Posterior Sampling}
	
	The Spherical Gaussian Constrained Posterior Sampling (PS SG) technique addresses the limitations of previous methods, namely standard PS and PS MC. Both earlier methods rely on the strong assumption that the data manifold $\mathcal{M}_0$ is linear and that the Jensen gap is negligible. However, as demonstrated in \cite{yang2024guidance}, this assumption is flawed, with the Jensen approximation introducing errors in the guidance term. The study in \cite{yang2024guidance} highlights that these errors are strongly dependent on the dimensionality of the problem, as indicated by a lower bound.
	
	To mitigate these issues, PS SG introduces a spherical Gaussian constraint to the guidance term. This constraint ensures that the guidance update remains within a high-confidence interval for $x_t$ to lie on the intermediate manifold $\mathcal{M}_t$, as defined for unconditional diffusion models. The high-confidence interval is represented as the hypersphere $\mathbb{S}^n_{\mu_\theta(x_t, c, t), \sqrt{n}\sigma_t}$, where $n$ is the dimensionality of the generated images. This modification refines the update rule, which is now expressed as:
	
	\begin{equation}
		x_{t-1} = \mu_\theta(x_t, c, t) - \sqrt{n}\sigma_t \frac{\nabla_{x_t}\|c - S_\phi(x_{0|t})\|^2_2}{\|\nabla_{x_t}\|c - S_\phi(x_{0|t})\|^2_2\|_2}.
	\end{equation}
	
\begin{figure}[h]
		\centering
		\includegraphics[scale=0.5]{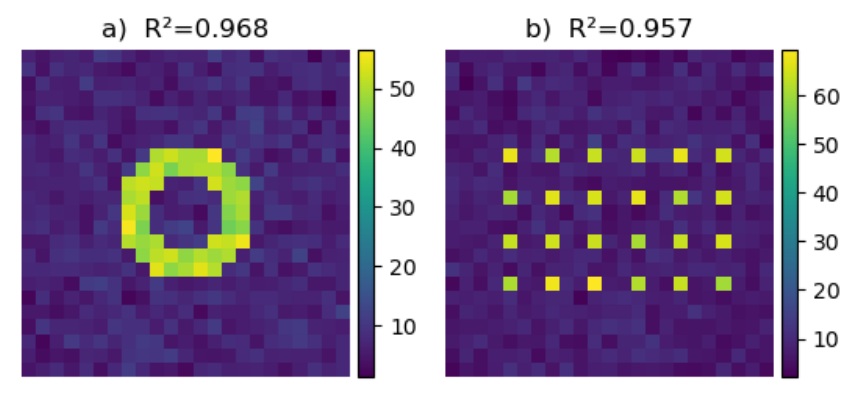}
		\caption{$|$Far field$|$ after simulation from inverse-designed metasurface parameters using Spherical Gaussian constrained Posterior Sampling.}
		\label{fig:dps_sg_results}
	\end{figure}

	The detailed derivation of this update rule is provided in \cite{yang2024guidance}. The correction term introduced by the PS SG method further improves alignment with the target, outperforming previous posterior sampling methods and other inverse design methods. This is illustrated in Figure \ref{fig:dps_sg_results}, as well as in the comparative results shown in Table \ref{table}.
	
	\section{Comparative Results}\label{sec:comp_res}
	The results presented for both GD and DMs correspond to 1,000 steps, reflecting the optimized training configuration for the DMs.

	\begin{table}[h]
		\caption{Comparison of the best diffusion model-based method, specifically PS SG, with the reference method using GD, evaluated using the $R^2$ metric. On average, PS SG outperforms GD and Phase Retrieval \& Look-Up Table approach.}
		\centering
		\begin{tabular}{|l|l|l|}
			\hline
			Method & $mean(R^2)$  & $std(R^2)$\\
			\hline
			PS without normalized guidance & 0.941  & 0.005\\ 
			\hline
			PS with normalized guidance & 0.905  & 0.024 \\ 
			\hline
			PS MC$_5$ without normalized guidance & 0.932 & 0.013 \\ 
			\hline
			PS MC$_5$ with normalized guidance & 0.876 & 0.008 \\ 
			\hline
			\textbf{PS SG} & \textbf{0.962}  & \textbf{0.006}\\ 
			\hline
			Phase Retrieval \& Look-Up Table & 0.913  & 0.020 \\ 
			\hline
			Gradient Descent & 0.697 & 0.053  \\ 
			\hline
		\end{tabular}
		
		\label{table}
	\end{table}
	
	Qualitatively, Figure~\ref{fig:dps_sg_results} illustrates the superior performance of the PS SG method over other PS techniques. This observation is quantitatively supported by Table~\ref{table}, where PS SG demonstrates both the highest average performance relative to reference methods and great consistency, as indicated by its low standard deviation in the $R^2$ metric. For further comparison, the far fields of GD and Phase Retrieval \& Look-Up-Table are provided in Appendix~\ref{App:method}.
	
\begin{figure}[h]
		\centering
		\includegraphics[scale=0.68]{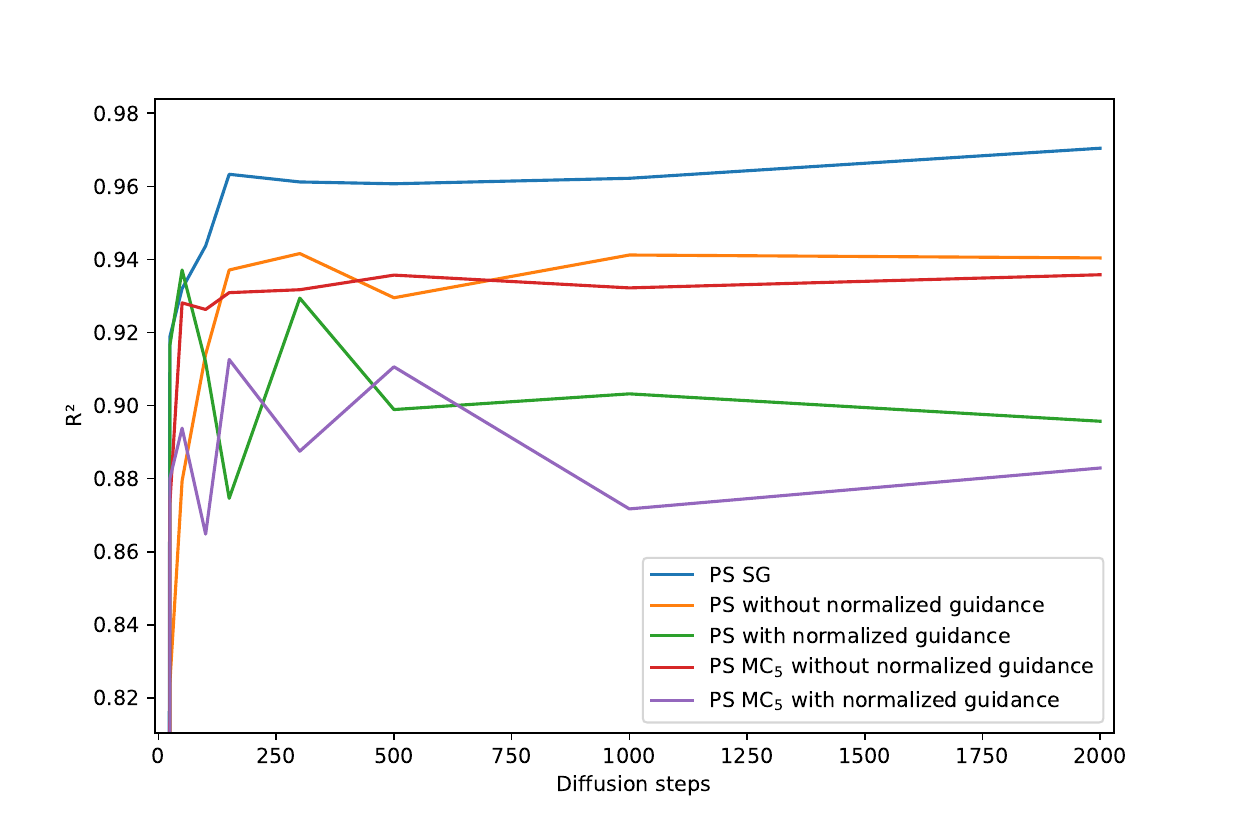}
		\caption{Comparison of $R^2$ values across varying numbers of sampling steps for the presented posterior sampling methods: PS, PS MC$_5$, and PS SG. Notably, PS SG consistently outperforms the other methods.}
		\label{fig:r2_steps_study}
	\end{figure}

An analysis of the number of function evaluations or diffusion steps, as depicted in Figure~\ref{fig:r2_steps_study}, indicates that Monte Carlo posterior sampling generally underperforms relative to other posterior sampling techniques. Beyond its slightly lower accuracy, this method demands greater computational resources due to the requirement for surrogate $S_\phi$ inference at each Monte Carlo sample.

Conversely, posterior sampling with Spherical Gaussian constraints consistently surpasses other techniques. It achieves faster accuracy improvement, reaches higher performance levels, and plateaus rapidly. This efficiency allows for a reduction in the number of diffusion steps, directly translating to shorter generation times while maintaining high precision. Consequently, decreasing the number of diffusion steps by a given factor yields a proportional reduction in overall generation time.

	\section{Scaling up} \label{sec:scaling}
	
	Due to memory and computational speed constraints of FDTD rigorous simulations, the database was limited to small metasurfaces of size $23 \times 23$ pillars, significantly smaller than the target size of approximately $100  \times  100$ pillars. This choice enabled the generation of a database containing 5k elements within two weeks of simulations, despite the ultimate goal of performing inverse design for metasurfaces of around $100 \times 100$ pillars. It is important to note that quantifying the precision of metasurfaces exceeding approximately $100 \times 100$ pillars presents challenges due to the substantial computational resources required for rigorous FDTD simulations. Consequently, the surrogate solver and DMs were trained on the $23 \times 23$ pillar database. 
	
\begin{figure}[h!]
		\centering
		\hspace{-1cm}
		\includegraphics[scale=0.7]{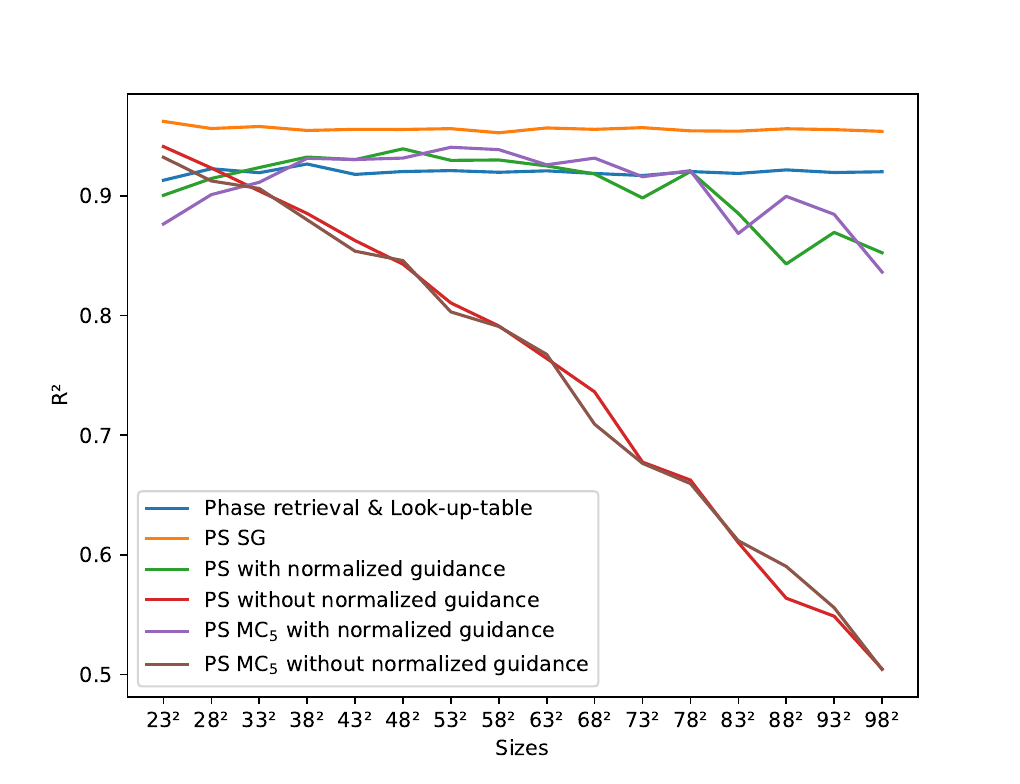}
		\caption{Comparison of $R^2$ values for the far field magnitude obtained from inverse-designed metasurfaces using GD, Phase Retrieval \& Look-Up Table, and DM methods, evaluated as the number of pillars increases. The GD method consistently underperforms, with its precision declining sharply with larger pillar arrays, leading to its exclusion from the plots due to low accuracy.The results highlight the critical role of guidance term normalization: it ensures that PS and PS MC maintain high precision, whereas their $R^2$ values degrade significantly without normalization as the number of pillars grows.}
		\label{fig:scaling_r2}
	\end{figure}

	However, the method demonstrates the capability to scale up to larger metasurfaces without requiring additional training. As illustrated in Figure \ref{fig:scaling_r2}, these models successfully scale to metasurfaces of larger sizes while maintaining a high level of precision. Using the DM PS SG method, metasurfaces comprising $98 \times 98$ pillars were generated in 2 minutes on an NVIDIA A100 GPU. $|$Far field$|$ for this larger metasurfaces can be found in Appendix \ref{App:scaling}.
	
\begin{figure}[h!]
		\centering
	
		\includegraphics[scale=0.55]{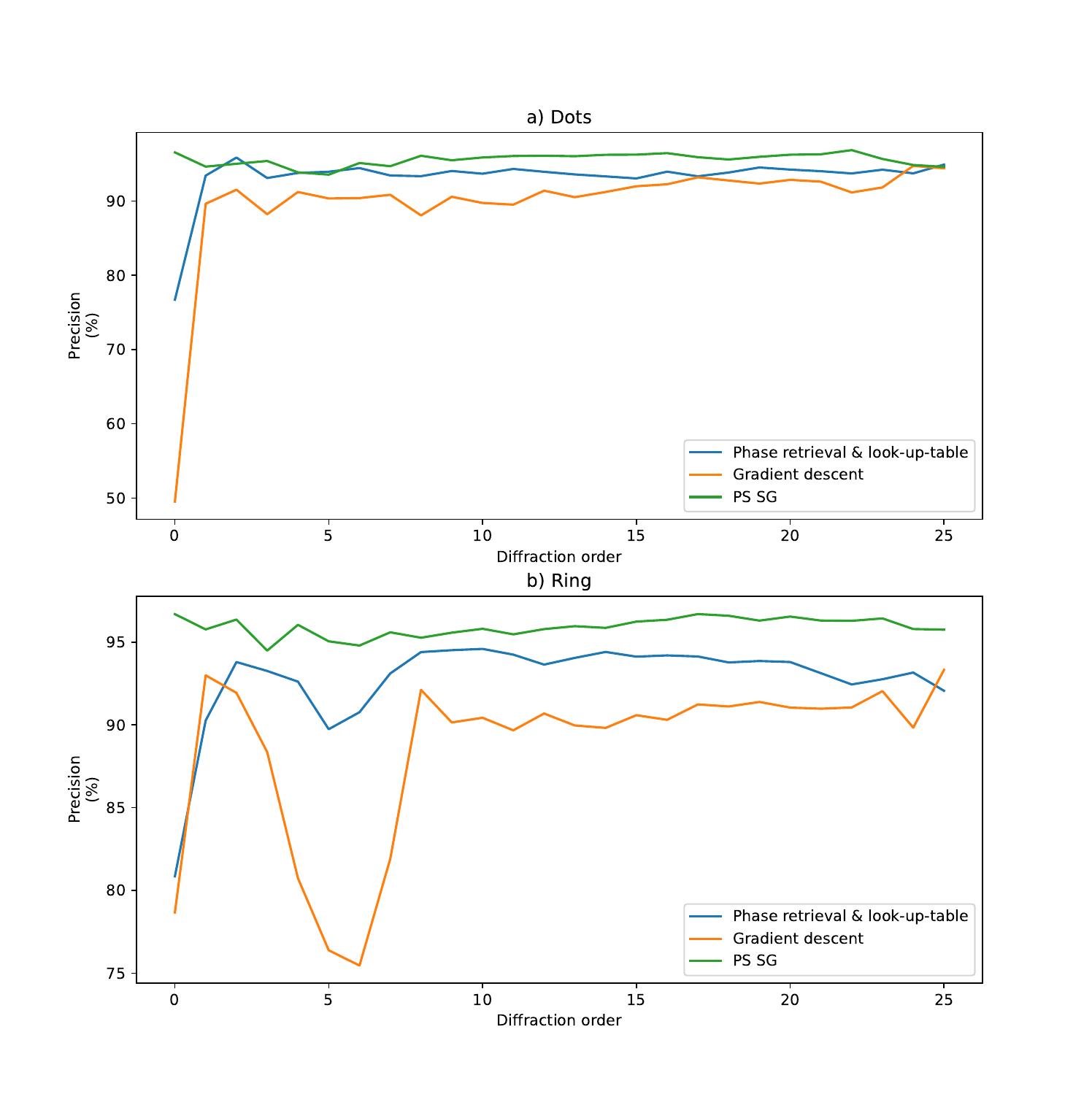}
		\caption{Precision performance across all diffraction orders for a $38 \times 38$ pillar metasurface designed using DM PS SG, Phase Retrieval \& Look-Up Table, and Gradient Descent. DM PS SG consistently outperforms the other methods, maintaining nearly constant precision across all orders.}
		\label{fig:scaling_ppp_38}
	\end{figure}

\begin{figure}[h!]
		\centering
	
		\includegraphics[scale=0.55]{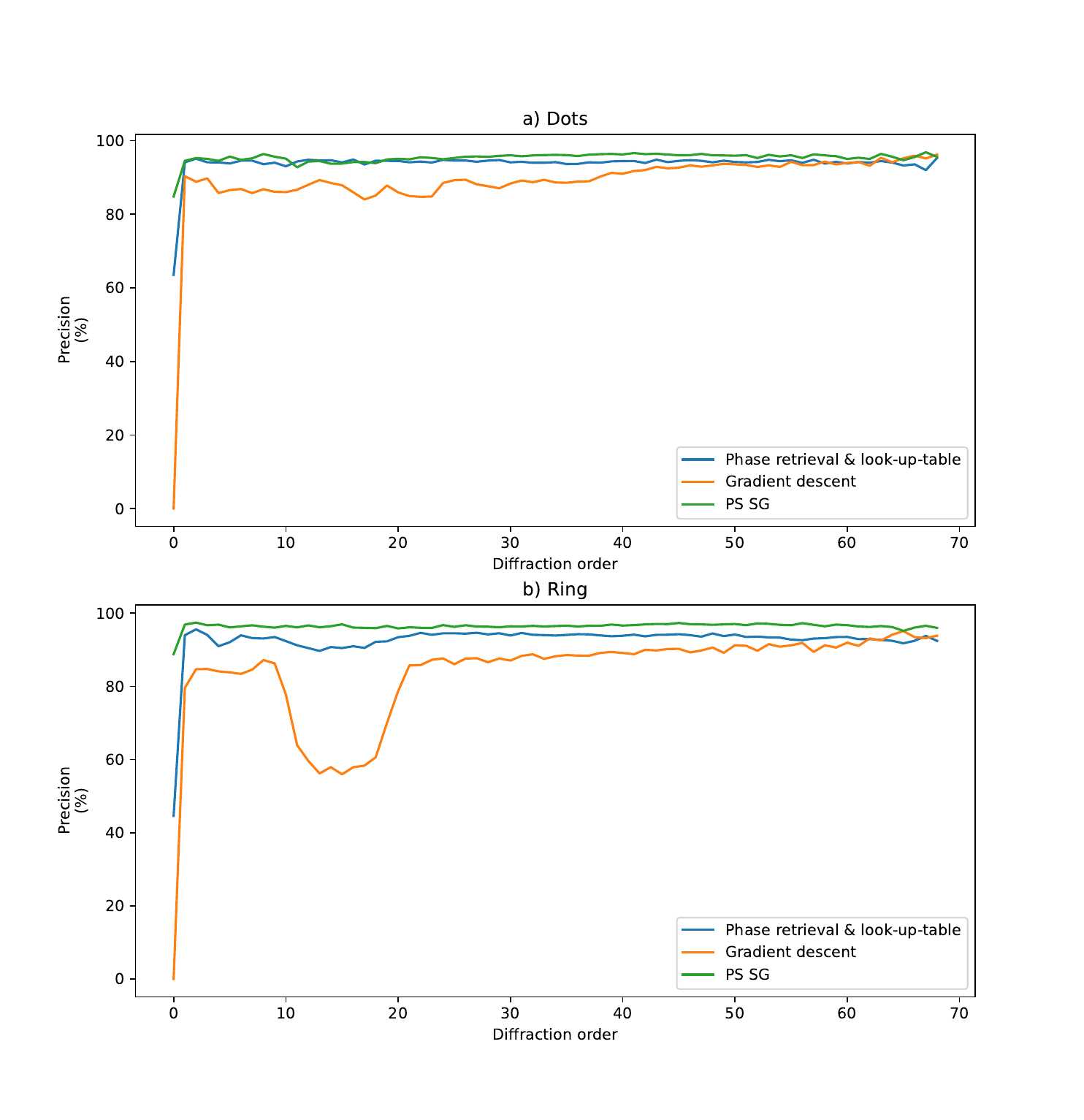}
		\caption{Precision performance across all diffraction orders for a $98 \times 98$ pillar metasurface designed using DM PS SG, Phase Retrieval \& Look-Up Table, and GD. DM PS SG consistently outperforms the other methods and $0^{th}$ dropping down like for other methods. }
		\label{fig:scaling_ppp_98}
	\end{figure}

	The precision of inverse design methods is rigorously evaluated by analyzing performance across individual diffraction orders, as illustrated in Figure~\ref{fig:scaling_ppp_38} and Figure~\ref{fig:scaling_ppp_98}. Among the PS methods assessed, PS SG consistently maintains high precision across all diffraction orders, with only a slight decline observed at the $98 \times 98$ pillar configuration for the $0^{th}$ order.
	
	In all cases, PS SG surpasses both Phase Retrieval \& Look-Up Table and GD approaches at every diffraction order. The latter two methods demonstrate significantly inferior performance, particularly for the $0^{th}$ diffraction order, with their accuracy further diminishing as the array size increases.
	The precision for diffraction order $n$ is defined as:
	$Precision_n = \frac{|Simulation_n - target_n|}{\max(target) - \min(target)}.$
	
	\FloatBarrier
	
	\section{Conclusion}
	
	Diffusion models employing basic ancestral sampling proved insufficient for far field metasurface inverse design. However, the incorporation of advanced posterior sampling techniques, including Monte Carlo and Spherical Gaussian constrained sampling, yielded state-of-the-art performance ($R^2= 0.962$). These methods significantly outperformed raw posterior sampling ($R^2 = 0.941$) and alternative approaches such as gradient descent ($R^2 = 0.697$) and phase retrieval with look-up tables ($R^2 = 0.913$) in terms of both accuracy and scalability.
	
	Unlike local approximate models, which face challenges with strong inter-meta-atom coupling or multilayer metasurfaces, diffusion models enhanced with posterior sampling reliably converge to global optima. This ensures superior precision and broad applicability across diverse metasurface configurations. Additionally, we introduced a scalable framework enabling inverse design models trained on small metasurfaces (23$\times$23 meta-atoms) to generalize effectively to significantly larger metasurfaces (98$\times$98 meta-atoms) without retraining. This approach maintains high accuracy and achieves rapid design generation in under three minutes, representing a substantial advancement in practical, high-fidelity metasurface inverse design.
	
\clearpage
\bibliography{references.bib}
\bibliographystyle{unsrturl}
\onecolumn
\appendix

\section{Noising schedule} \label{App:schedule}
Three distinct scheduling functions for the noise variance $\beta_t$ are introduced, with further options detailed in \cite{le_grand_2026_18148500}. These functions govern the noise addition schedule during the diffusion process as follows:

\begin{itemize}
	\item Linear schedule : $\beta_t = \beta_{\text{start}} + t(\beta_{\text{end}}-\beta_{\text{start}})$
	\item Quadratic schedule : $\beta_t = (\sqrt{\beta_{\text{start}}} + t(\sqrt{\beta_{\text{end}}}-\sqrt{\beta_{\text{start}}}))^2$
	\item Sigmoid schedule : $\beta_t = \beta_{\text{start}} + \frac{1}{1+e^{t}}(\beta_{\text{end}}-\beta_{\text{start}})$
\end{itemize}

\begin{figure}[h]
	\centering
	\includegraphics[scale=0.5]{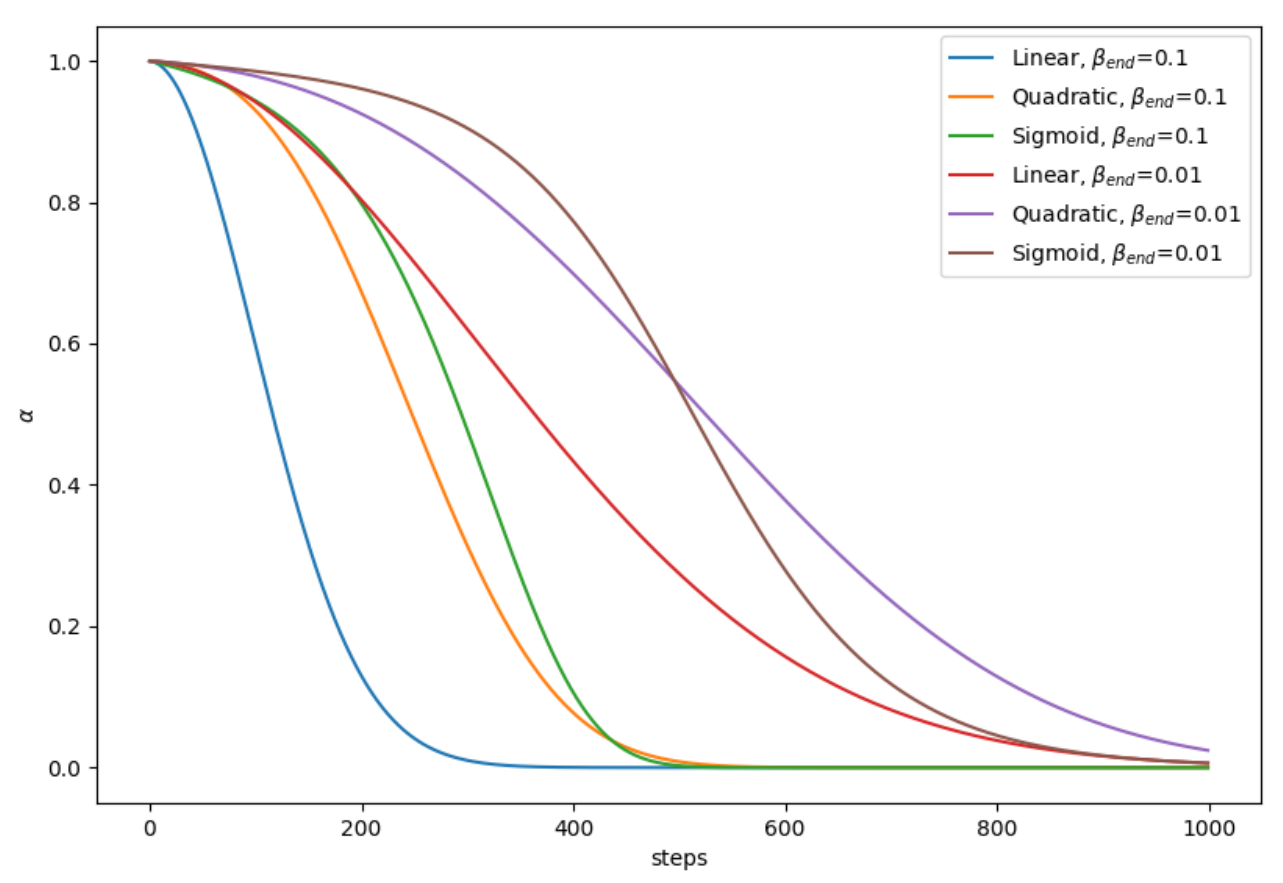}
	\caption{Evaluation of the data coefficient $\alpha(t)$, computed from $\beta(t)$ using Eq. \eqref{eq:alpha}, for three noise schedules linear, sigmoid, and quadratic each tested with $\beta_{\text{end}} \in \{0.1, 0.01\}$ while fixing $\beta_{\text{start}} = 10^{-4}$. The y-axis represents the post-generation performance metric $R^2_{far field, S_\phi}$, assessed on far field using the surrogate network $S_\phi$, and the x-axis denotes the training metric, which quantifies the diffusion network's accuracy in predicting the added noise ($\epsilon$).}
	\label{schedule_alpha}
\end{figure}

Different noising schedules are illustrated in Figure~\ref{schedule_alpha}, highlighting the substantial variability among the schedules.

\begin{figure}[h]
	\centering
	\includegraphics[scale=0.55]{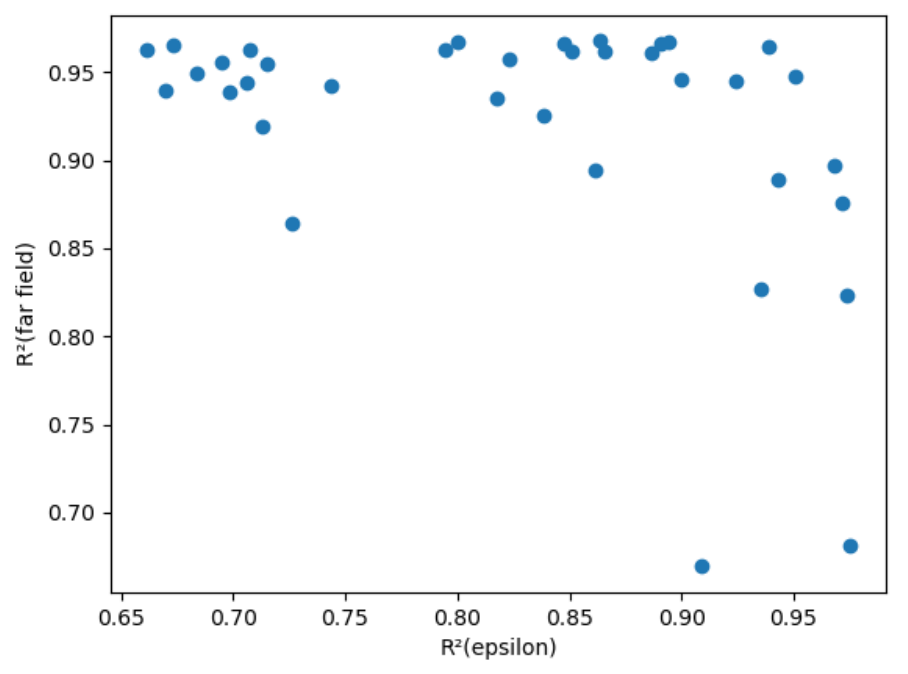}
	\caption{Each blue dot represents a diffusion model trained with distinct schedule parameters drawn from the set $\{ \text{schedule function} \} \times \{ \beta_{\text{end}} \} \times \{ \text{consistency loss} \} = \{ \text{sigmoid}, \text{quadratic}, \text{linear} \} \times \{ 0.1, 0.01, 0.001 \} \times \{ \text{no consistency}, \text{scheduled consistency}, \text{consistency} \}$. When combined with enhanced sampling techniques, the best training metric does not necessarily correspond to the best final metric.}
	\label{schedule_results}
\end{figure}

The noise schedule significantly impacts the final performance, a relationship not immediately evident from training metrics alone. As shown in Figure~\ref{schedule_results}, two key observations emerge:

First, the post-generation surrogate-simulation performance, measured by $R^2_{far field, S_\phi}$, exhibits significant variability. While certain configurations achieve state-of-the-art results ($R^2_{far field, S_\phi} \approx 0.99$), others perform below established benchmarks ($R^2_{far field, S_\phi} \approx 0.65$).

Second, there is no observable correlation between training metrics and final evaluation performance. Consequently, a comprehensive verification process encompassing full generation and comparison against the specified conditions is essential for evaluating diffusion models in metasurface inverse design.

\section{Surrogate network}\label{App:surrogate}

Similar to the diffusion network, the objective was to train the surrogate model on a database of small-sized metasurface elements and generalize its performance to larger configurations during inference. To ensure scalability, a fully convolutional network architecture was adopted \cite{long2015fully}, incorporating skip connections and residual blocks to facilitate the training of deeper networks \cite{ronneberger2015u,he2016deep}. The selected architecture comprises approximately 775k parameters, representing an optimal balance: deeper or larger networks did not yield performance improvements, while smaller networks resulted in degraded accuracy, given the available computational resources and data.

An additional consideration for the network size, is the computational overhead during generation, where gradient computation from the surrogate network consumes significant RAM. This competes with the memory allocated for metasurface parameters, ultimately impacting the scaling capability the maximum size of metasurfaces that can be generated which is critical for addressing the problem effectively.

To maintain spatial consistency under periodic boundary conditions, periodic padding was applied to all convolutional layers. The surrogate model's scalability was evaluated using simulations with pillar configurations ranging from $28^2$ to $98^2$. The model achieved an average precision on test data characterized by an $R^2$ value of 97.73 and a standard deviation of $\sigma_{R^2} = 0.07$ across the tested range.

\section{Method comparison} \label{App:method}

\begin{figure}[h]
	\centering
	\includegraphics[scale=0.4]{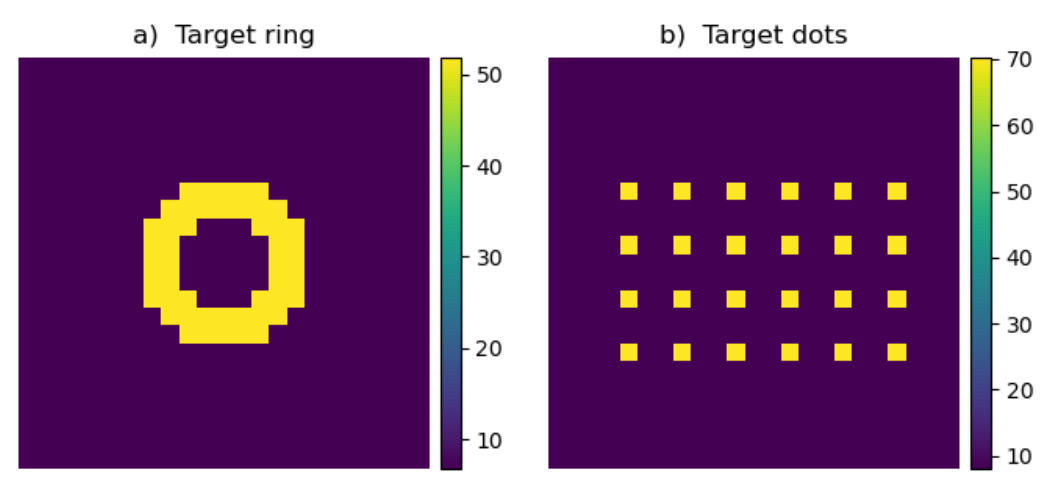}
	\caption{$|$Far field$|$ targets (a, b) used as reference patterns throughout the paper.}
	\label{fig:targets}
\end{figure}

The far field condition provided as input to both the diffusion model and the surrogate network is represented as a two-value image, as illustrated in Figure~\ref{fig:targets}. The assigned values are determined by the spatial distribution of light concentration, while the total energy remains consistent across both images in Figure~\ref{fig:targets}.

\begin{figure}[h]
	\centering
	\includegraphics[scale=0.4]{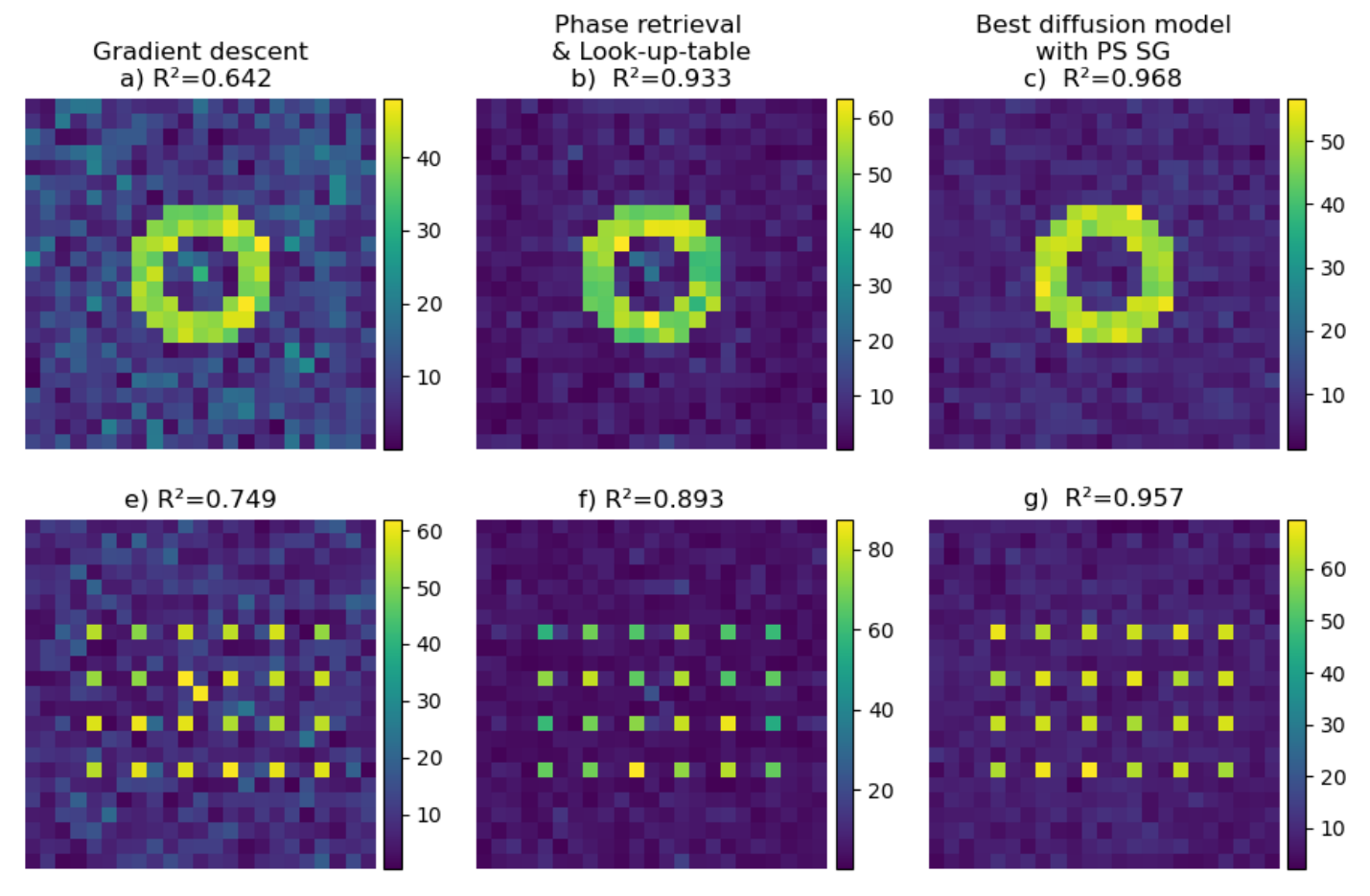}
	\caption{$|$Far field$|$ after simulation from inverse-designed metasurface parameters using, Gradient Descent (a,e), Phase Retrieval \& Look-Up Table (b, f) and the best Diffusion Model (c, g) with PS SG.}
	\label{fig:methods_comparison}
\end{figure}

Figure~\ref{fig:methods_comparison} presents a comparative analysis of the image results obtained using Gradient Descent, Phase Retrieval \& Look-Up-Table, and the Diffusion Model with Posterior Sampling incorporating Spherical Gaussian constraints (the highest-performing PS technique). In the results from Gradient Descent and Phase Retrieval \& Look-Up-Table, an unwanted $0^{th}$ diffraction order is prominently visible for $23 \times 23$ metasurfaces. In contrast, this artifact is entirely absent in the results generated by the Diffusion Model with Posterior Sampling and Spherical Gaussian constraints.

\section{Scaling results in image} \label{App:scaling}

\begin{figure}[h!]
	\centering
	\includegraphics[scale=0.35]{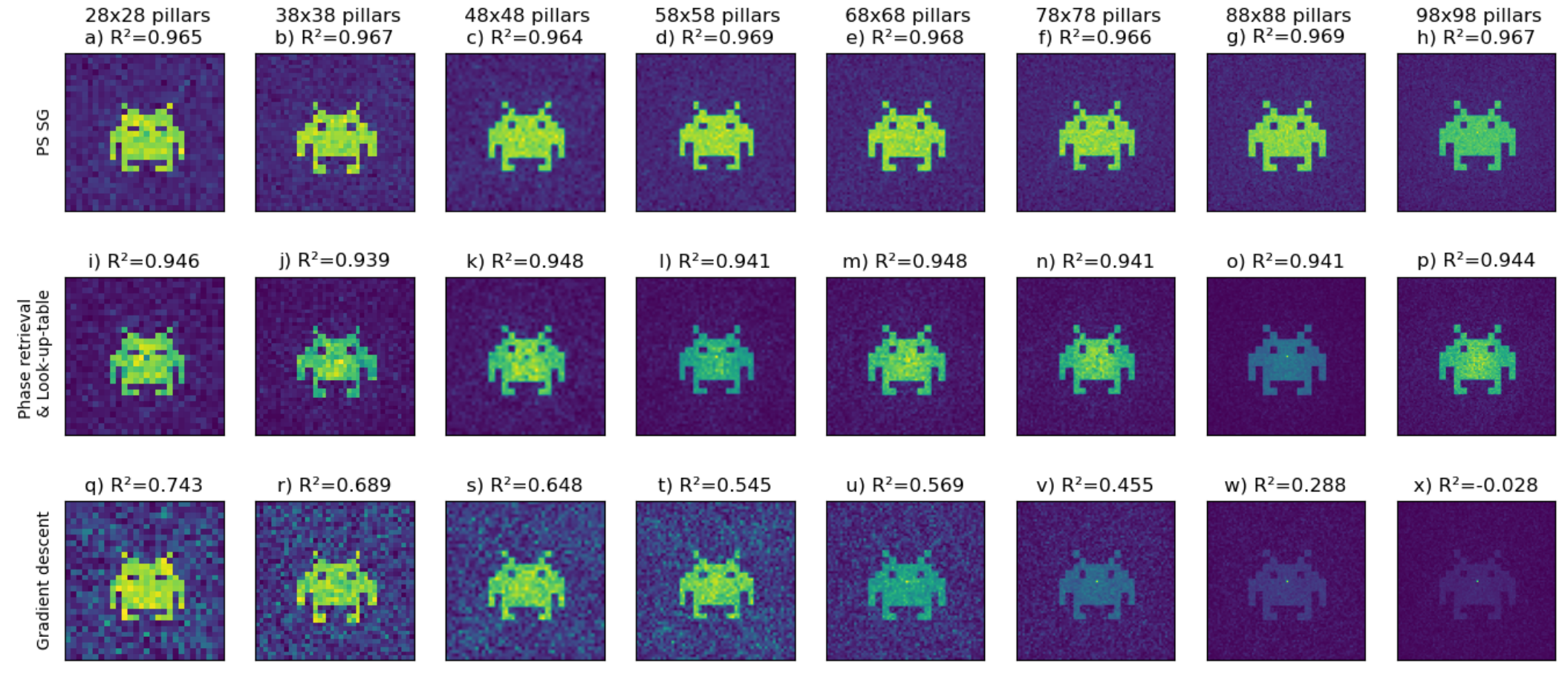}
	\caption{Far field results obtained from FDTD simulations of a space invader design using Gradient Descent, Phase Retrieval \& Look-Up Table, and Diffusion Models (DMs) for metasurfaces of increasing size, ranging from $28^2$ to $98^2$ pillars.}
	\label{fig:scaling_image_space_invader1}
\end{figure}

\begin{figure}[h!]
	\centering
	\includegraphics[scale=0.35]{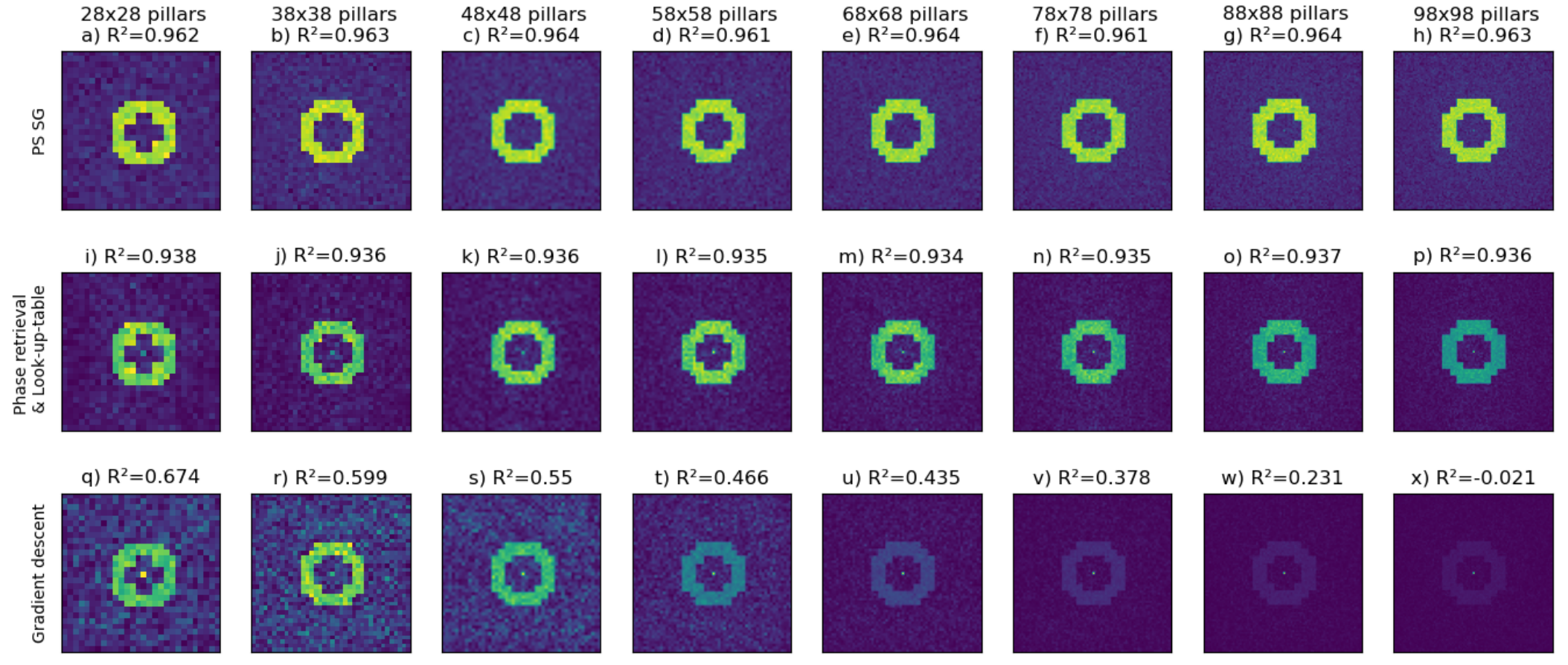}
	\caption{Far field results obtained from FDTD simulations of a ring design using Gradient Descent, Phase Retrieval \& Look-Up Table, and Diffusion Models (DMs) for metasurfaces of increasing size, ranging from $28^2$ to $98^2$ pillars.}
	\label{fig:scaling_image_ring1}
\end{figure}

\begin{figure}[h!]
	\centering
	\includegraphics[scale=0.35]{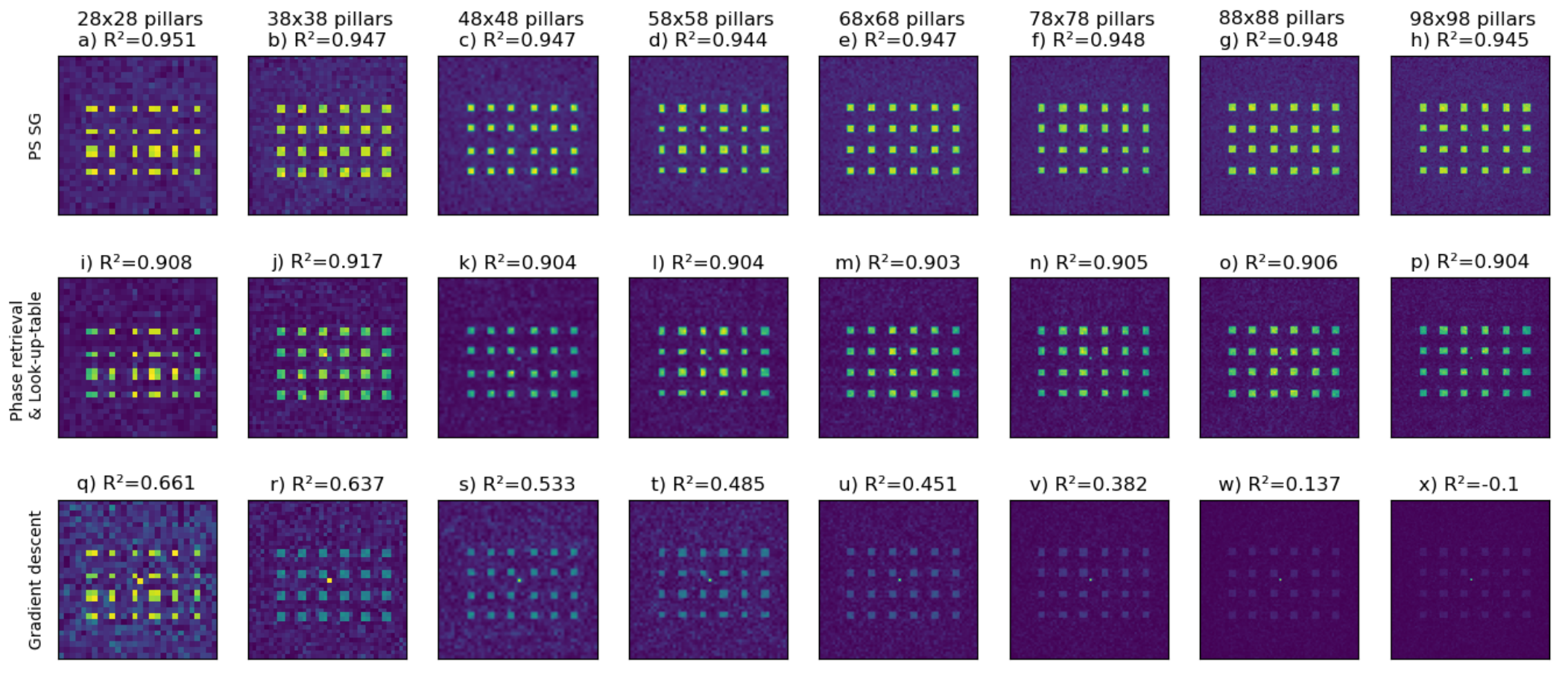}
	\caption{Far field results obtained from FDTD simulations of dots design using Gradient Descent, Phase Retrieval \& Look-Up Table, and Diffusion Models (DMs) for metasurfaces of increasing size, ranging from $28^2$ to $98^2$ pillars.}
	\label{fig:scaling_image_dots1}
\end{figure}

In Figures~\ref{fig:scaling_image_space_invader1},~\ref{fig:scaling_image_ring1}, and~\ref{fig:scaling_image_dots1}, the degradation of the far field pattern relative to the target is evident as the metasurface size increases, particularly for the Gradient Descent method. For this method, the target pattern becomes barely discernible for metasurfaces exceeding $78 \times 78$ pillars.

Additionally, the Diffusion Model with Posterior Sampling and Spherical Gaussian (PS SG) demonstrates a more uniform energy distribution across all target shapes compared to other methods. In contrast, the Phase Retrieval \& Look-Up-Table method tends to concentrate energy toward the center of the shape.

\end{document}